\documentclass[10pt,twoside]{article}

\usepackage{epsfig}
\usepackage{graphicx}
\usepackage{rotating}
\usepackage{fleqn}
\usepackage{espcrc2}

\newcommand{\pvec}{{\bf p}}
\def\etal {{\it et~al.,}}
\input babarsym

\title{\begin{flushleft}\small{Presented at QCD 02: High Energy Physics International Conference
       in Quantum Chromodynamics, Montpellier, France, 2-9 Jul 2002.}\\
       \vskip 5pt
       \small{Nucl. Phys. B (Proc. Suppl.) 121 (2003) 239-248}\\
       \small{SLAC-PUB-10138}
       \end{flushleft}         
       \vskip 20pt
       $B$ Decays at \babar}

\author{J.J.Back\address{Physics Department, Queen Mary, University of London, \\
	Mile End Road, London, E1 4NS, UK 
	(on behalf of the \babar\ Collaboration)}%
	\thanks{Work supported in part by the Department of Energy
	contract DE-AC03-76SF00515.}}

\begin{document}

\begin{abstract}
We present branching fraction and \CP\ asymmetry 
results for a variety of
$B$ decays based on up to 56.4~\invfb collected by 
the \babar\ experiment running near the \FourS resonance at 
the \pep2 \epem $B$-factory.
\end{abstract}

\maketitle

\section{The \babar\ Detector}

The results presented in this paper are based on an integrated luminosity
of up to 56.4~\invfb collected at the \FourS\ resonance with the 
\babar\ detector~\cite{babardet}
at the \pep2\ asymmetric \epem\ collider at the Stanford Linear Accelerator
Center. Charged particle track parameters are measured by a 
five-layer double-sided silicon vertex tracker and a 40-layer drift chamber
located in a 1.5-T magnetic field. Charged particle identification is
achieved with an internally reflecting ring imaging Cherenkov detector
(DIRC) and from the average \dedx energy loss measured in the tracking devices. Photons
and $\piz$s are detected with an electromagnetic calorimeter (EMC) consisting
of 6580 CsI(Tl) crystals. An instrumented flux return (IFR), containing multiple
layers of resistive plate chambers, provides muon and long-lived hadron
identification.

\section{$B$ Decay Reconstruction}
\label{sec:BDecayReco}

The $B$ meson candidates are identified kinematically using
two independent variables. The first is $\Delta E = E^{*} - E^{*}_{beam}$, which is
peaked at zero for signal events, since the energy of the $B$ candidate in the
\FourS\ rest frame, $E^{*}$, must be equal to the energy of the beam, $E^{*}_{beam}$,
by energy conservation.
The second is the beam-energy substituted mass,
$\mes = \sqrt{(E^{*2}_{beam} - \pvec^{*2}_B)}$, where $\pvec^{*}_B$
is the momentum of the $B$ meson in the \FourS\ rest frame, and must be close to
the nominal $B$ mass~\cite{pdg}. 
The resolution of $\mes$ is dominated by the beam energy
spread and is approximately 2.5~\mevcc.

Several of the $B$ modes presented here have decays that involve 
neutral pions ($\pi^0$) and $\KS$ particles. Neutral pion candidates are
formed by combining pairs of photons in the EMC, with requirements made to the
energies of the photons and the mass and energy of the $\pi^0$. 
Table~\ref{tab:pi0andKscuts}
shows these requirements for various decay modes, as well as 
the selection requirements for $\KS$ candidates, which are
made by combining oppositely charged pions.
\begin{table*}[!htb]
\begin{center}
\caption{Selection requirements for $\pi^0$ and $\KS$ candidates for 
various $B$ decay modes ($h = K/\pi$). 
$E_{\gamma}$ is the minimum photon energy and
$m_{\piz}$ and $E_{\piz}$ the mass and energy, respectively, of $\piz$ candidates.
The mass of the $\KS$ is $m_{\KS}$, the opening angle between the $\KS$ momentum
and its line-of-flight is $\phi_{\KS}$, the transverse
flight distance of the $\KS$ from the primary event vertex is $d_{\KS}$ and
$\tau/\sigma_{\KS}$ is the $\KS$ lifetime divided by its error.}
\label{tab:pi0andKscuts}
{\footnotesize
\begin{tabular}{lccccccc}
\hline
Mode & $E_{\gamma}~(\mev)$ & $m_{\pi^0}~(\mevcc)$ & $E_{\pi^0}~(\mev)$ 
& $m_{\KS}~(\mevcc)$ & $\phi_{\KS}$~(mrad) & $d_{\KS}$~(mm) & $\tau/\sigma_{\KS}$ \\

\hline
$B \ra DK$ & $>70$ & $[124,144]$ & $>200$ & --- & --- & --- & --- \\
$B \ra D^{(*)} D^{(*)}$ & $>30$ & $[115,155]$ & $>200$ & $[473,523]$ & $<200$ & $>2$ & ---\\
$B \ra h \pi^0$ & $>30$ & $[111,159]$ & --- & --- & --- & --- & ---\\
$B \ra h K^0$ & --- & --- & --- & $[487,509]$ & --- & --- & $>5$ \\
$B \ra \phi K^{(*)}$ & --- & --- & --- & $[487,510]$ & $<100$ & --- & $>3$ \\
$B \ra \eta h$ & $>50$ & $[120,150]$ & --- & --- & --- & --- & --- \\
$B \ra \eta K^0$ & $>50$ & $[115,155]$ & --- & $[491,507]$ & $<40$ & $> 2$ & --- \\
$B \ra \eta' K^{(*)0}$ & --- & --- & --- & $[488,508]$ & --- & --- & --- \\
$B \ra K^{*} \gamma$ & $>30$ & $[115,150]$ & $>200$ & $[489,507]$ & --- & --- & --- \\
$B \ra K^{(*)} \ell^+ \ell^-$ & --- & --- & --- & $[480,498]$ & --- & $>1$ & --- \\
\hline
\end{tabular}}
\end{center}
\end{table*}

Significant backgrounds from light quark-antiquark continuum events
are suppressed using various event 
shape variables which exploit the difference in the event topologies
in the centre-of-mass frame between
background events, which have a di-jet structure, and signal events, 
which tend to be rather spherical. One example
is the cosine of the angle $\theta^*_T$ between the thrust axis of the signal 
$B$ candidate and the thrust axis of the 
rest of the tracks and neutrals in the event.
This variable is strongly peaked at unity for 
continuum backgrounds and is flat for signal.
\begin{figure}[hbt!]
\begin{center}
\vspace*{-0.5cm}
\epsfig{file=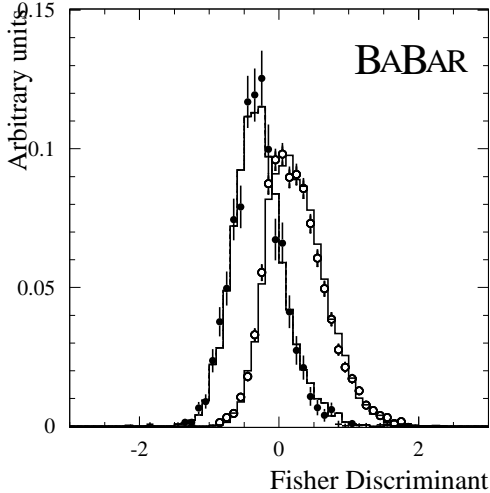,width=0.45\textwidth,clip=}
\caption{The Fisher distribution for $B^0 \ra \pi^+\pi^-$ signal 
Monte Carlo simulated events (left histogram)
compared to data $B^- \ra D^0 \pi^-$ decays (black points), and
continuum background Monte Carlo (right histogram) compared to on-resonance
sideband data (open points).}
\label{fig:Fisher}
\end{center}
\end{figure}

Further suppression of backgrounds can be achieved using
a Fisher discriminant ${\cal{F}}$, which is a linear 
combination of event shape variables, such as the scalar sum of the
centre-of-mass momenta of all charged tracks and neutrals, excluding the
$B$ decay products, flowing into nine concentric cones centred on the
thrust axis of the $B$ candidate. Signal events have a lower Fisher
discriminant value compared to background events, as shown in Fig.~\ref{fig:Fisher}.

Sidebands in on-resonance (\FourS) data are used to characterise the light quark 
background
in $\Delta E$ and $\mes$, as well as data taken at 40~\mev below the \FourS\
resonance (``off-resonance''). The phenomenologically motivated Argus 
function~\cite{Argus} is used to fit the background $\mes$ distributions. 
Control samples are used to compare
the performance between Monte Carlo simulated events and on-resonance data.

The results presented here are from either
extended maximum likelihood fits or from cut-and-count methods.
All of the analyses have been performed ``blind'', meaning that
the signal region is looked at only after the selection criteria
have been finalised (in order to reduce the risk of bias).

Charge conjugate modes are implied
throughout this paper.

\section{$B \ra D K$}

\begin{figure}[htb!]
\begin{center}
\epsfig{file=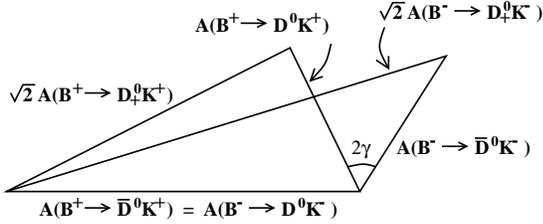,width=0.45\textwidth,clip=}
\caption{Relation between the amplitudes for the processes $B \ra D K$
and the angle $\gamma$.}
\label{fig:gammafig}
\end{center}
\end{figure}

The decays $B^- \ra D^0 K^-$ and 
$B^{\pm} \ra D^0_{\pm} K^{\pm}$, where $D^0_{\pm}$
denotes the \CP-even ($+$) or \CP-odd states ($-$)
$(D^0 \pm \bar{D}^0)/\sqrt{2}$,
can be used to measure in a theoretically clean way 
the Cabibbo-Kobayashi-Maskawa (CKM) angle $\gamma$, one
of the parameters that describes \CP\ Violation in the 
Standard Model~\cite{gamma1}~\cite{gamma2}.
Figure~\ref{fig:gammafig} shows a graphical representation of the relation
between the amplitudes of the decays and the angle $\gamma$. 

This measurement is experimentally challenging because the branching fractions
of these decays are of the order of $10^{-7}$. However, in the meantime we can measure
the ratio between the branching fractions for $B^- \ra D^0 K^-$
and $B^- \ra D^0 \pi^-$, where the final states for the $D^0$ meson
are $K^-\pi^+$, $K^- \pi^+ \pi^- \pi^+$ and $K^- \pi^+ \pi^0$. 
The invariant masses of $D^0 \ra K^- \pi^+, K^- \pi^+ \pi^- \pi^+$ 
candidates are required to be 
within $20.4~\mevcc$ ($3~\sigma$) of the 
nominal value~\cite{pdg}. Since the mode $D^0 \ra K^- \pi^+ \pi^0$
has more combinatorial background compared to the other decays,
the invariant mass is required to 
be within two standard deviations ($2 \times 11~\mevcc$). 

Three event shape variables are used to suppress light-quark continuum background
events. The first is the normalised Fox-Wolfram moment $H_2/H_0$~\cite{FoxWolfram},
which is required to be less than 0.5 for all selected events. The second variable
is $\theta^*_T$, mentioned in section~\ref{sec:BDecayReco}. The value
of $|$cos$\theta^*_T|$ is required to be less than 0.9 for the $D^0 \ra K^+ \pi^-$
mode and less than 0.7 for $D^0 \ra K^- \pi^+ \pi^- \pi^+$ and $D^0 \ra K^- \pi^+ \pi^0$.
Thirdly, we use the helicity angle $\theta_{hel}$, defined as the angle
between the direction of the $D^0$ candidate calculated in the rest frame
of the $B$ and the direction of one of the decay products of the $D^0$, calculated
in the rest frame of the $D^0$. The distribution of cos$\theta_{hel}$ is flat
for signal and peaked at $\pm 1$ for fake $D^0$ events. For continuum events,
$|$cos$\theta^*_T|$ and cos$\theta_{hel}$ are correlated, and we use
\begin{eqnarray}
|\mbox{cos}\theta_{hel}| < 0.9; & 0.0 \leq |\mbox{cos}\theta^*_T| < 0.7 \nonumber \\
|\mbox{cos}\theta_{hel}| < -3|\mbox{cos}\theta^*_T| + 3; & 0.7 \leq |\mbox{cos}\theta^*_T| \leq 1.0. \nonumber \\
\end{eqnarray}

The yields for the signal modes are found by using an unbinned extended 
maximum likelihood fit
to the $\Delta E$ and $\mes$ variables, together with the Cherenkov angles
$(\theta_C$) of the prompt tracks measured by the DIRC (to distinguish kaons
from pions). We obtain
\begin{eqnarray}
R & = & \frac{\BR(B^- \ra D^0 K^-)}{\BR(B^- \ra D^0 \pi^-)} \nonumber \\
& = & (8.31 \pm 0.35 \pm 0.13)\%,
\label{DKREqn}
\end{eqnarray}
where the first error is statistical and the second error is systematic.
This quantity has also been measured by the CLEO and BELLE Collaborations,
where they get $R = (5.5 \pm 1.4 \pm 0.5)$\%~\cite{CLEODKR} and
$R = (7.9 \pm 0.9 \pm 0.6)$\%~\cite{BELLEDKR}, respectively. Theory predicts, using
factorisation and tree-level Feynman diagrams only, 
a value $R \approx \mbox{tan}^2\theta_C(f_K/f_{\pi})^2 \approx 7.4$\%, where
$\theta_C$ is the Cabibbo angle, and $f_K$ and $f_{\pi}$ are the meson
decay constants.
For the \CP-even mode $D^0_+ \ra K^+ K^-$ we have measured
\begin{eqnarray}
R_{CP} & = &\frac{\BR(B^- \ra D^0_+ K^-) + \BR(B^+ \ra D^0_+ K^+)}
{\BR(B^- \ra D^0_+ \pi^-) + \BR(B^+ \ra D^0_+ \pi^+)} \nonumber \\
& = & (8.4 \pm 2.0 \pm 0.8)\%,
\label{DKCPREqn}
\end{eqnarray}
and the direct \CP\ asymmetry
\begin{eqnarray}
{\cal{A}}_{CP} & = & \frac{\BR(B^- \ra D^0_+ K^-) - \BR(B^+ \ra D^0_+ K^+)}
{\BR(B^- \ra D^0_+ K^-) + \BR(B^+ \ra D^0_+ K^+)} \nonumber \\
& = &0.15 \pm 0.24 ^{+0.07}_{-0.08}.
\label{DKCP}
\end{eqnarray}

\section{$B \ra D^{(*)} D^{(*)}$}

Time-dependent \CP\-violating asymmetries in the 
decays $B \ra D^{(*)}D^{(*)}$ can be used to measure the CKM angle
$\beta$~\cite{Dstar1}, in a way complimentary to measurements already
made with decays such as $B^0 \ra J/\psi \KS$~\cite{JpsiKs}.
However, the vector-vector decay of $B^0 \ra D^{*+} D^{*-}$ is not a
pure \CP\ eigenstate, which may cause a sizeable dilution to 
the \CP\ violation that can be observed. In principle, a full time-dependent
angular analysis can remove this dilution~\cite{Dstar2}.

We reconstruct exclusively the decays $B^0 \ra D^{*+} D^{*-}$ and
$B^0 \ra D^{*\pm}D^{\mp}$, where $D^{*\pm} \ra D^0 \pi^{\pm}$ or $D^{\pm} \pi^0$.
The final states we consider for the neutral $D$ mesons are $K^-\pi^+$,
$K^- \pi^+ \pi^0$, $K^- \pi^+ \pi^- \pi^+$ and $\KS \pi^+ \pi^-$, while
we consider the $D^+$ final states $K^- \pi^+ \pi^+$, $\KS \pi^+$ and $K^-K^+\pi^+$.

$B^0$ candidates are reconstructed by performing a mass-constrained
fit to the $D$ and $D^*$ candidates. In the case when more than one $B$ candidate
is found for an event, we chose the $B$ candidate in which the 
$D$ and $D^*$ mesons have invariant masses closest to their nominal values~\cite{pdg}.

Signal events are required to satisfy $|\Delta E| < 25~\mev$ and 
$5.273 < \mes < 5.285~\gevcc$.

Using a sample of 22.7 million \BB\ pairs, we obtain
the following branching fractions 
\begin{equation}
\BR(B^0 \ra D^{*+} D^{*-}) = (8.0 \pm 1.6 \pm 1.2) \times 10^{-4},
\label{DStarBR1}
\end{equation}
\begin{equation}
\BR(B^0 \ra D^{*\pm} D^{\mp}) = (6.7^{+2.0}_{-1.7} \pm 1.1) \times 10^{-4}.
\label{DstarBR2}
\end{equation}
The fraction of the \CP-odd component $R_t$ of $B^0 \ra D^{*+} D^{*-}$ decays
can be found by using the angular distribution of the decay in the transversity
basis~\cite{Dstar2}
\begin{equation}
\frac{1}{\Gamma} \frac{d\Gamma}{d{\mbox{cos}}\theta_{tr}}
= \frac{3}{4}(1 - R_t){\mbox{sin}}^2\theta_{tr} + 
\frac{3}{2}R_t{\mbox{cos}}^2\theta_{tr},
\end{equation}
where $\Gamma$ is the decay rate and $\theta_{tr}$ is the polar angle between
the normal to the $D^{*-}$ decay plane and the $\pi^+$ line of flight 
in the $D^{*+}$ rest frame. Using an unbinned extended maximum likelihood fit, 
we find
\begin{equation}
R_t = 0.22 \pm 0.18 \pm 0.03.
\label{Rt}
\end{equation}

\section{Two-body Charmless $B$ Decays}
Measurements of the branching fractions and \CP\ asymmetries for 
$B$ decays into two-body charmless final states will give us important information
about the CKM angles $\alpha$ and $\gamma$~\cite{twobody}. 
The time-dependent \CP-violating
asymmetry in the decay $B^0 \ra \pi^- \pi^+$ is related to the
angle $\alpha$. If the decay proceeds only through tree diagrams, then the asymmetry
is directly related to $\alpha$. However, we can only measure 
an effective angle, $\alpha_{eff}$, if there is pollution from gluonic penguins.
In principle, $\alpha$ can be extracted by using the isospin-related 
decays $B^+ \ra \pi^+ \pi^0$ and $B^0 \ra \pi^0 \pi^0$~\cite{alpha}.
It is also possible to get a bound on the value of $\alpha$~\cite{alphaBound} via
\begin{equation}
\mbox{sin}^2(\alpha_{eff} - \alpha) 
< \frac{\left<\BR(B^0 \ra \pi^0\pi^0)\right>_{CP}}{\BR(B^+ \ra \pi^+ \pi^0)},
\label{alphaBoundEqn}
\end{equation}
where 
$\left<\BR(B^0 \ra \pi^0\pi^0)\right>_{CP} = 
\frac{1}{2} [ \BR(B^0 \ra \pi^0 \pi^0) +$ \\
$\BR(\bar{B}^0 \ra \pi^0 \pi^0)]$.
The angle $\gamma$ can be constrained by using ratios of
branching fractions for various $\pi\pi$ and $K\pi$ decays~\cite{gamma3}.

Here, we consider only decays involving at least one neutral
particle in the final state
(see~\cite{hhref} for results on $B^0 \ra h^+ h^-$ decays, where
$h = \pi$ or $K$). 
\begin{figure}[htb!]
\begin{center}
\epsfig{file=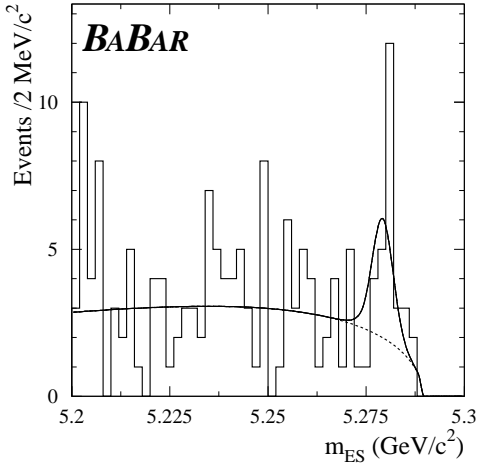,width=0.9\linewidth,clip=}
\caption{Distribution of $\mes$ for $B^+ \ra \pi^+ \pi^0$ events in on-resonance data. 
The solid curve 
represents the projection of the maximum likelihood fit result, while
the dotted curve represents the background contribution.}
\label{fig:pipi0mes}
\end{center}
\end{figure}

Backgrounds from non-hadronic events 
are reduced by requiring 
$H_2/H_0$~\cite{FoxWolfram} to be less than 0.95 and the sphericity~\cite{sphericity} 
of the event to be greater than 0.01.
Light-quark continuum events are suppressed by using two other event shape variables.
The first is the angle $\theta^*_S$ in the centre-of-mass frame 
between the sphericity axis of the $B$
candidate and of the remaining tracks and neutrals in the event. The distribution
of $|$cos$\theta^*_S|$ is peaked near unity for background and is 
approximately uniform for signal events. We require $|$cos$\theta_S|$ to be less
than 0.9 for $h\KS$, 0.8 for $h \pi^0$ and 0.7 for $\pi^0\pi^0$. 

The second variable is the Fisher discriminant ${\cal{F}}$ mentioned in 
section~\ref{sec:BDecayReco}.

The branching fractions and charge asymmetries shown
in Table~\ref{tab:twobodyresults} are obtained from an 
unbinned extended maximum likelihood fit using ${\cal{F}}$, $\theta_C$,
$\mes$ and $\Delta E$. Figure~\ref{fig:pipi0mes} shows the
$\mes$ distribution for $B^+ \ra \pip\piz$ decays, in which we observe a signal
for the first time. 
\begin{table}[!htb]
\begin{center}
\caption{Two-body charmless $B$ decay branching fractions ($\BR$) and $\CP$
asymmetries (${\cal{A}}_{CP}$) based on $56.4~\invfb$. Upper limits are given
at the 90\% confidence level.}
\label{tab:twobodyresults}
{\footnotesize
\begin{tabular}{lccc}
\hline
Mode & $\BR$ ($10^{-6}$) & ${\cal{A}}_{CP}$ \\
\hline
$\pip\piz$ & $4.1^{+1.1 +0.8}_{-1.0 -0.7}$ & 
$-0.02^{+0.27}_{-0.26} \pm 0.10$\\
$\Kp\piz$ & $11.1^{+1.3}_{-1.2} \pm 1.0$ & $0.00 \pm 0.11 \pm 0.02$ \\
$\pip\Kz$ & $17.5^{+1.8}_{-1.7} \pm 1.8$ & $-0.17 \pm 0.10 \pm 0.02$ \\
$\Kp\Kzb$ & $< 1.4 (-0.6 ^{+0.6}_{-0.7} \pm 0.3)$ & --- \\
$\piz\piz$ & $< 3.4 (-0.9 ^{+0.9 + 0.8}_{-0.7 -0.6})$ & --- \\
\hline
\end{tabular}}
\end{center}
\end{table}

\section{Three-body Charmless Charged $B$ \\Decays}

The decays $B^+ \ra h^+ h^- h^+$, where $h$ = $\pi$ or $K$, can
be used to measure the angle $\gamma$~\cite{B3body}. The basic idea is that
there can be interference between resonant and non-resonant amplitudes 
leading to direct \CP\ violation. A Dalitz plot analysis can, in principle, 
give us information about all of the strong and weak phases in these decays.
A first step towards this goal is to measure the branching fractions into the whole
Dalitz plot. We can write these as
\begin{equation}
\BR = \frac{1}{N_{\BB}} \sum_i \frac{S_i}{\epsilon_i},
\label{B3bodyEqn}
\end{equation}
where $N_{\BB}$ is the total number of $\BB$ pairs in the data sample,
$S_i$ is the net signal (after background subtraction) in cell $i$ of the Dalitz plot
and $\epsilon_i$ is the signal efficiency in that cell 
found from Monte Carlo simulation.
No assumptions are made about intermediate resonances.

$B$ candidates are formed by combining three charged tracks. 
We use $\dedx$ information from the tracking devices and the Cherenkov
angle and number of photons measured by the DIRC for tracks with momenta above
$700~\mevc$, to identify charged pions and kaons. Electron candidates
are vetoed by requiring that they fail a selection based on information from 
$\dedx$, shower shapes in the EMC and the ratio of the shower energy and 
track momentum.
We remove $B$ candidates when the combination of any two of its (oppositely
charged) daughter tracks is within $3~\sigma$ of the mass of the 
$D^0$, $J/\psi$ or $\psi(2S)$ mesons. Here, $\sigma$ is $10.0~\mevcc$ for 
$D^0$ and $15.0~\mevcc$ for $J/\psi$ and $\psi(2S)$.

Continuum backgrounds are suppressed by requiring selections on $|$cos$\theta^*_T|$
and on the Fisher discriminant mentioned in section~\ref{sec:BDecayReco}.
Comparisons between Monte Carlo simulated events 
and on-resonance data are made using
the control sample $B^- \ra D^0 (\ra K^- \pi^+) \pi^-$, which has a similar
decay topology as the charmless signal modes.

The signal region is defined as $|\mes - m_B| < 8.0~\mevcc$ and
$|\Delta E - \left<\DeltaE\right>| < 60.0~\mev$, where
$\left<\Delta E\right>$ is $7.0~\mev$ (obtained from the $D^0\pi^-$ control 
sample) and $m_B$ is the nominal mass of the charged $B$ meson~\cite{pdg}.
\begin{figure}[htb!]
\begin{center}
\epsfig{file=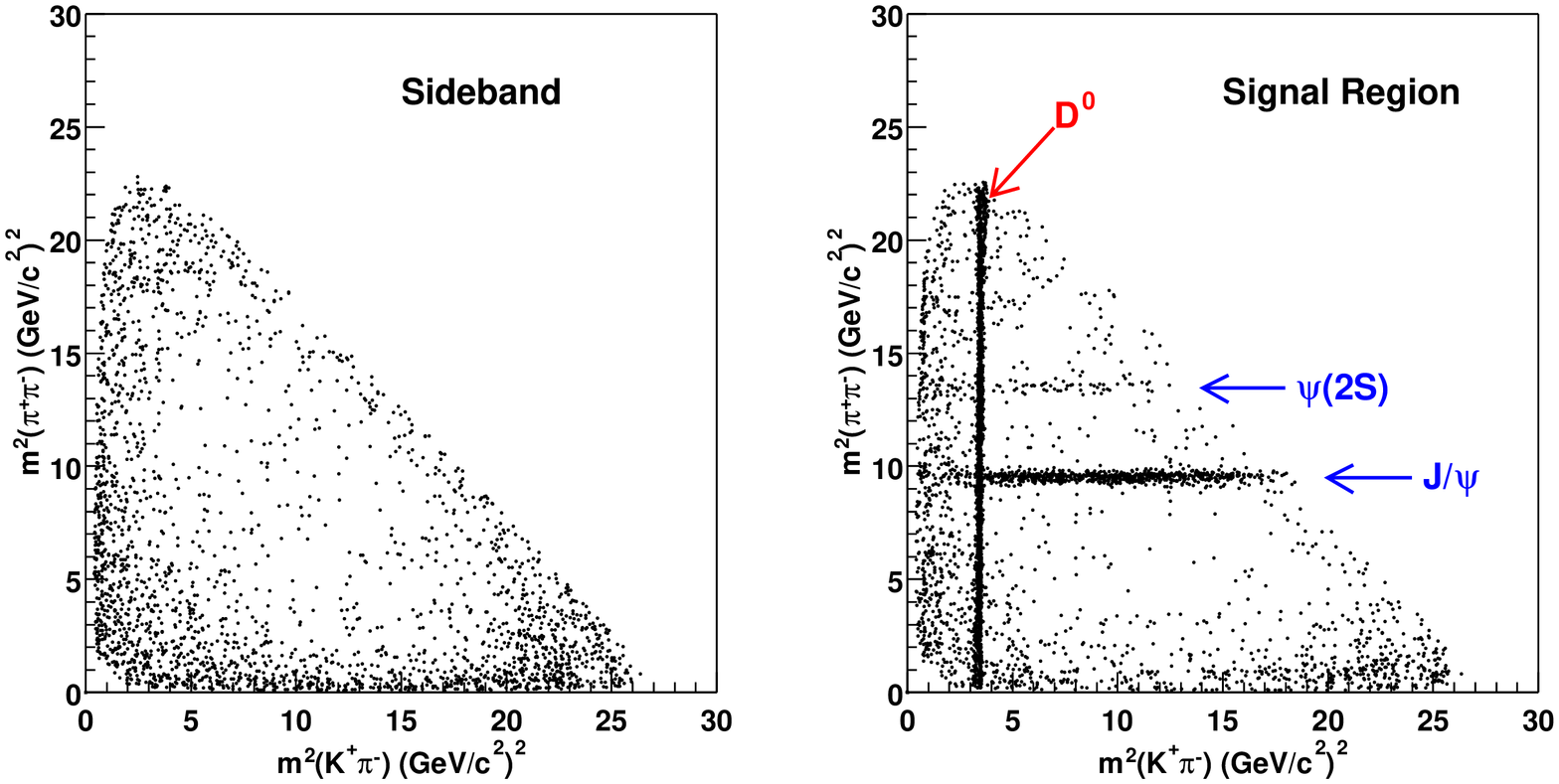,width=0.6\linewidth,angle=-90,clip=}\hfill
\epsfig{file=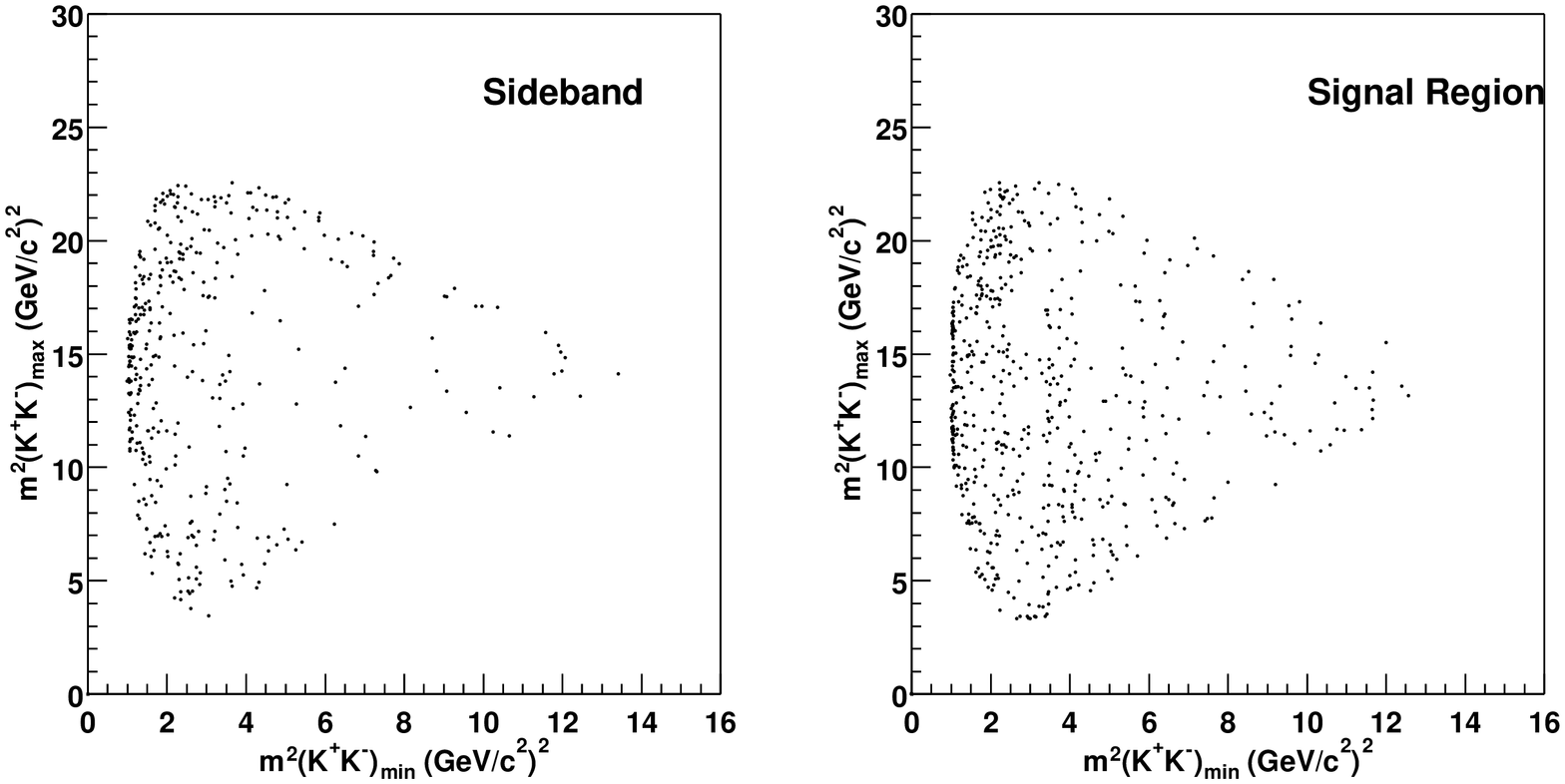,width=0.6\linewidth,angle=-90,clip=}
\caption{Unbinned Dalitz plots (with no background subtraction or efficiency
corrections) for $B^+ \ra K^+ \pi^- \pi^+$ events
in on-resonance sideband (top left) and signal (top right) data, and for
$B^+ \ra K^+ K^- K^+$ events in on-resonance sideband (bottom left) 
and signal (bottom right) data. Open charm contributions are not removed.}
\label{fig:B3bodyDP}
\end{center}
\end{figure}

Table~\ref{tab:B3bodyTable} shows the results for $51.5~\invfb$, 
where we have also included the results from the BELLE Collaboration~\cite{B3bodyBELLE} 
for comparison. Figure~\ref{fig:B3bodyDP} shows preliminary
unbinned Dalitz plots for $B^+ \ra K^+ \pi^- \pi^+$ and
$B^+ \ra K^+ K^- K^+$ in on-resonance data (with no background subtraction or efficiency
corrections applied).
As a cross-check, we measure $(180 \pm 4 \pm 11) \times 10^{-6}$
for the branching fraction for the $B^- \ra D^0 \pi^-$ control sample, which
agrees with the previously measured value of 
$(203 \pm 20) \times 10^{-6}$~\cite{pdg}.
\begin{table}[!htb]
\begin{center}
\caption{Three-body charmless charged $B$ decay branching fractions 
($\times 10^{-6}$) from \babar\ ($51.5~\invfb$) and BELLE ($29.1~\invfb$).}
\label{tab:B3bodyTable}
{\footnotesize
\begin{tabular}{lcc}
\hline
Mode &  $\babar$ & BELLE\\
\hline
$\pi^+\pi^-\pi^+$ & $8.5 \pm 4.0 \pm 3.6 (< 15)$ & --- \\
$K^+\pi^-\pi^+$ & $59.2 \pm 4.7 \pm 4.9$ & $55.6 \pm 5.8 \pm 7.7$\\
$K^+K^-\pi^+$ & $2.1 \pm 2.9 \pm 2.0 (< 7)$ & $< 21$ \\
$K^+K^-K^+$ & $34.7 \pm 2.0 \pm 1.8$ & $35.3 \pm 3.7 \pm 4.5$ \\
\hline
\end{tabular}}
\end{center}
\end{table}

\section{$B \ra \phi K^{(*)}, \phi \pi$}
\label{sec:phiKsec}

These modes are interesting because only penguin diagrams contribute
to the decay amplitudes (mainly $b \ra s \bar{s} s$),
and the time-dependent \CP\ asymmetry for the
neutral mode $B^0 \ra \phi \KS$ 
can be used to measure sin$2\beta$. Comparison with sin$2\beta$ results from
charmonium modes will allow us to probe new physics participating in penguin
loops~\cite{Dstar1}. Isospin symmetry predicts that 
$\BR(B^{\pm} \ra \phi K^{\pm}) \approx \BR(B^0 \ra \phi K^0)$,
and there are theoretical estimates for the various branching fractions
based on factorisation models and perturbative QCD~\cite{phiKQCD}.

$B$ mesons are reconstructed by combining $\phi \ra K^+ K^-$ candidates
with either a $K^*$ candidate or a bachelor charged track. We consider
the decays $K^{*+} \ra K^0 \pi^+$ (with $\KS \ra \pi^+ \pi^-$), 
$K^{*+} \ra K^+ \pi^0$ and $K^{*0} \ra K^+ \pi^-$. 

Continuum backgrounds are suppressed by using a Fisher discriminant and requiring that 
$|$cos$\theta^*_T| < 0.9$.

Table~\ref{tab:phiKresults} shows the branching fractions for the modes 
obtained from an extended unbinned maximum likelihood fit using
$\mes$, $\Delta E$, $\theta_C$, the mass of the $\phi$ resonance, the Fisher discriminant
and the cosine of the helicity angle. 
Also shown in Table~\ref{tab:phiKresults} are the results
from the BELLE~\cite{phiKBELLE} and CLEO~\cite{phiKCLEO} collaborations.
\begin{table*}
\begin{center}
\caption{Branching fractions ($\times 10^{-6}$) for $B \ra \phi K$ 
and $B^+ \ra \phi \pi^+$ decays
from \babar\ ($56.3~\invfb$ and $20.7~\invfb \dagger$), BELLE ($21.6~\invfb$)
and CLEO ($9.1~\invfb$).}
\label{tab:phiKresults}
\begin{tabular}{lccc}
\hline
Mode & $\babar$ & BELLE & CLEO \\
\hline
$\phi \Kp$ & $9.2 \pm 1.0 \pm 0.8$ & $11.2^{+2.2}_{-2.0} \pm 1.4$ & $5.5^{+2.1}_{-1.8} \pm 0.6$\\
$\phi \Kz$ & $8.7^{+1.7}_{-1.5} \pm 0.9$ & $8.9^{+3.4}_{-2.7} \pm 1.0$ & $< 12.3$\\
$\phi K^{*+}$ & $9.7^{+4.2}_{-3.4} \pm 1.7 \dagger$ & $< 36$ & $< 22.5$ \\
$\phi K^{*0}$ & $8.7^{+2.5}_{-2.1} \pm 1.1 \dagger$ & $13.0^{+6.4}_{-5.2} \pm 2.1$ & $11.5^{+4.5 + 1.8}_{-3.7 - 1.7}$\\
$\phi \pip$ & $< 0.6$ & --- & $< 5$\\
\hline
\end{tabular}
\end{center}
\end{table*}

\section{Charmless $B$ decays with $\eta$ and $\eta'$ mesons}

These rare decays proceed via $b \ra u$ tree and $b \ra s$
penguin Feynman diagrams. Interference between the various amplitudes can give
rise to direct \CP\ violation and the time-dependent \CP\ asymmetries for the neutral
modes are sensitive to the value of sin$2\beta$. The CLEO collaboration
has observed unexpectedly high branching fractions for $B \ra \eta K^{*}$
and $B \ra \eta' K$~\cite{etaCLEO}, leading some theorists to
speculate on exotic processes such as QCD anomalies and penguins with an 
enhanced charm contribution in the virtual loop~\cite{etaTheory}.

We have studied the decays $B \ra \eta h, \eta K^*, \eta' K$ and $\eta' K^{*0}$.
The $\eta$ resonances are formed by combining two photons, each with a minimum energy
of $50~\mev$, while $\eta'$ mesons are reconstructed in the final states
$\eta' \ra \eta \pi^+ \pi^-$ or $\eta' \ra \rho^0 \gamma$, where $\rho^0 \ra \pi^+ \pi^-$.
The $\rho^0$ candidates are required to have an
invariant mass between 500 and $995~\mevcc$. We consider the same 
neutral and charged $K^*$ decays as those in the $B \ra \phi K^*$ analysis
mentioned in section~\ref{sec:phiKsec}.

Like other analyses, the requirement $|$cos$\theta^*_T| < 0.9$ is imposed
to suppress continuum background. 

We use the energy constrained mass $m_{EC}$, which is the mass of the $B$ candidate
when its energy is constrained to be equal to the beam energy, instead of $\mes$.

Table~\ref{tab:etaResults} shows the branching fraction results from \babar\
using extended unbinned maximum likelihood fits to the data. 
The main variables in the fits are $\m_{EC}$,
$\Delta E$, the invariant mass and helicity
distributions of the intermediate resonance and a Fisher
discriminant. 
Also shown are the results from CLEO~\cite{etaCLEO}
and BELLE. We confirm the large branching fractions for the
$\eta K^*$ and $\eta'K$ modes.
\begin{table*}[!htb]
\begin{center}
\caption{Measured branching fractions ($\times 10^{-6}$) for $B$ decays with $\eta$ and $\eta'$
mesons from the CLEO, BELLE and \babar\ collaborations.}
\label{tab:etaResults}
\begin{tabular}{lcccc}
\hline
\vspace*{0.3cm}
Mode & CLEO ($9.1\invfb$) & BELLE ($29\invfb$) & $\babar$ ($56\invfb$, $21\invfb$$\dagger$)\\
\hline
$\eta \pip$ & $1.2^{+2.8}_{-1.2} (< 5.7)$ & $5.4^{+2.0}_{-1.7}\pm 0.6$ & $2.2^{+1.8}_{-1.6}\pm 0.1 (< 5.2) \dagger$\\
$\eta \Kp$ & $2.2^{+2.8}_{-2.2} (< 6.9)$ & $5.3^{+1.8}_{-1.5}\pm 0.6$ & $3.8^{+1.8}_{-1.5}\pm 0.2 (< 6.4) \dagger$ \\
$\eta \Kz$ & $0.0^{+3.2}_{-0.0} (< 9.3)$ & & $6.0^{+3.8}_{-2.9}\pm 0.4 (< 12) \dagger$ \\
$\eta K^{*0}$ & $13.8^{+5.5}_{-4.6} \pm 1.6$ & $16.5^{+4.6}_{-4.2} \pm 1.2$ &$19.8^{+6.5}_{-5.6} \pm 1.5 \dagger$ \\
$\eta K^{*+}$ & $26.4^{+9.6}_{-8.2} \pm 3.3$ & $26.5^{+7.8}_{-7.0} \pm 3.0$ & $22.1^{+11.1}_{-9.2} \pm 3.2 \dagger$ \\
$\etapr \Kp$ & $80^{+10}_{-9} \pm 7$ & $78 \pm 6 \pm 9$ & $67 \pm 5 \pm 5$\\
$\etapr \Kz$ & $89^{+18}_{-16} \pm 9$ & $68 \pm 10$ & $46 \pm 6 \pm 4$\\
$\etapr K^{*0}$ & $7.8^{+7.7}_{-5.7} (< 24)$ & & $4.0^{+3.5}_{-2.4} \pm 1.0 (< 13)$\\
\hline
\end{tabular}
\end{center}
\end{table*}

\section{$B \ra K^* \gamma$}

The exclusive decays $B \ra K^* \gamma$ proceed via the flavour-changing
neutral $b \ra s \gamma$ transition, where the largest
contribution comes from the top quark in the electromagnetic penguin virtual loop. 
The current Standard Model next-to-leading order predictions for the 
branching fractions for these modes lies
between $3.5\times 10^{-4}$ and $6.2\times 10^{-4}$~\cite{Kstargamma}.
New physics contributions may enhance the observed branching fractions.

The decays $B^0 \ra K^{*0} \gamma$ and $B^+ \ra K^{*+} \gamma$
have been studied. The $K^{*}$ is formed by combining $K^+$, $\KS$,
$\pi^-$ and $\pi^0$ candidates through the four decay modes 
$K^{*0} \ra K^+ \pi^-, \KS \pi^0$ and 
$K^{*+} \ra K^+\pi^0, \KS \pi^+$. 

The background from these decays is predominantly from light-quark continuum
events, and are suppressed by requiring selections on $|$cos~$\theta^*_T|$ 
and the helicity angle.
The branching fractions are found by using an unbinned maximum likelihood
fit to the $\mes$ distributions, with the requirements
$-200 < \Delta E < 100~\mev$ for the $K^+\pi^-$ and $\KS \pi^+$ modes
and $-225 < \Delta E < 125~\mev$ for the modes containing
a $\pi^0$ ($K^+\pi^0$ and $\KS \pi^0$). Table~\ref{tab:KstarGamma} shows
the branching fraction and \CP\ asymmetry results for $20.7~\invfb$, where the
\CP\ asymmetry for these decays is defined as 
\begin{equation}
{\cal{A}}_{CP} = 
\frac{\BR (\bar{B} \ra \bar{K}^* \gamma) -
\BR (B \ra K^* \gamma)}
{\BR (\bar{B} \ra \bar{K}^* \gamma) +
\BR (B \ra K^* \gamma)}.
\label{KstarGammaAsym}
\end{equation}
Theoretical expectations for the branching fractions are in agreement
with the measured values.
\begin{table}
\begin{center}
\caption{Branching fraction ($\BR$) and \CP\ asymmetry (${\cal{A}}_{\CP}$) 
results for the $B \ra K^* \gamma$ channels for $20.7~\invfb$.}
\label{tab:KstarGamma}
{\footnotesize
\begin{tabular}{lcc} 
\hline
Mode                    &  $\BR (\times 10^{-5})$     & ${\cal{A}}_{\CP}$       \\
$B^0 \to K^{*0} \gamma$ & $4.23\pm 0.40\pm0.22$ & $-0.05 \pm 0.09 \pm 0.01$  \\ 
$B^+ \to K^{*+} \gamma$ & $3.83\pm 0.62\pm0.22$ & $-0.04 \pm 0.13 \pm 0.01$  \\
\hline
\end{tabular}}
\end{center}
\end{table}

\section{$B \ra K^{(*)} \ell^+ \ell^-$}

The decays $B \ra K^{(*)} \ell^+ \ell^-$, where $\ell^{\pm}$ is a charged
lepton, proceed via flavour-changing neutral currents, which are highly
suppressed in the Standard Model. The dominant contributions
come from one-loop electroweak penguins, with branching fractions predicted 
at the $10^{-7} - 10^{-6}$ level~\cite{KllTheory}. These could be enhanced
if new, heavy particles, such as those from supersymmetric
models, appear in the virtual loop. 

\begin{figure}[htb!]
\begin{center}
\epsfig{file=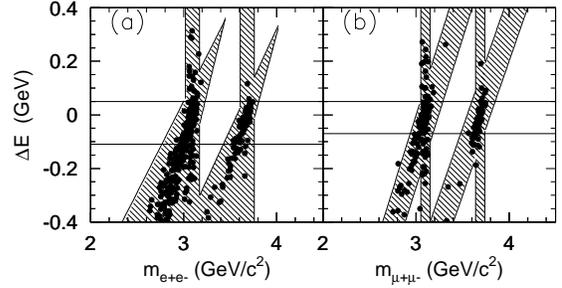,width=0.45\textwidth,clip=}
\caption{Veto in the $\Delta E$ vs.$m_{\ell^+\ell^-}$ plane for
(a) $B \ra K^{(*)} e^+ e^-$ and
(b) $B \ra K^{(*)} \mu^+ \mu^-$. The hatched regions
are vetoed. The dots correspond to 
$B \ra J/\psi(\ell^+ \ell^-) K$ and
$B \ra \psi(2S)(\ell^+ \ell^-) K$ 
Monte Carlo simulated events, while the
horizontal lines show the boundaries for the $\Delta E$ region where most
of the signal events would lie.}
\label{fig:KllFig}
\end{center}
\end{figure}

\begin{figure}[htb!]
\begin{center}
\epsfig{file=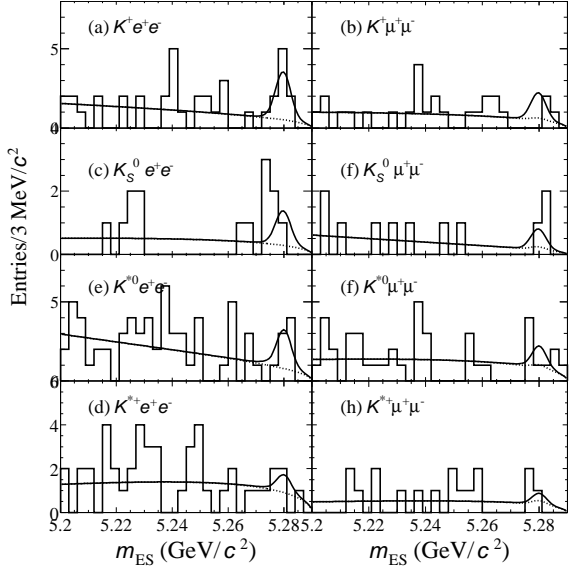,width=0.5\textwidth,clip=}
\caption{Projections from the likelihood fits of the
$K^{(*)} \ell^+ \ell^-$ modes onto $\mes$ for
the signal region $-0.11 < \Delta E < 0.05~\gev$ for electrons
and $-0.07 < \Delta E < 0.05~\gev$ for muons. The dotted lines
show the level of background, while the solid lines
show the sum of the signal and background contributions.}
\label{fig:KllPlots}
\end{center}
\end{figure}

We consider the four decay modes $B^+ \ra K^+ \ell^+ \ell^-$,
$B^0 \ra \KS \ell^+ \ell^-$, $B^+ \ra K^{*+} \ell^+ \ell^-$ 
and $B^0 \ra K^{*0} \ell^+ \ell^-$, where $K^{*0} \ra K^+ \pi^-$,
$K^{*+} \ra \KS \pi^+$, $\KS \ra \pi^+ \pi^-$, and $\ell$
is either an $e$ or $\mu$. We require the two oppositely charged leptons
to each have a momentum greater than 0.5 (1.0)$~\gev$ for $e (\mu)$.
Electron-positron pairs consistent with photon conversions are removed
from the data sample. 
$K^{*}$ candidates are required to have an invariant mass
within $75~\mevcc$ of the mean mass of $892~\mevcc$~\cite{pdg}.
The charm decays $B \ra J/\psi(\ell^+ \ell^-) K^{(*)}$ and
$B \ra \psi(2S)(\ell^+ \ell^-) K^{(*)}$ have identical topologies
to signal events, and are suppressed by applying a veto in the $\Delta E$
versus invariant mass of the lepton pair ($\m_{\ell^+\ell^-}$) plane,
as shown in Fig.~\ref{fig:KllFig}.
The signal is extracted using a two-dimensional extended unbinned maximum
likelihood fit to $\mes$ and $\Delta E$ in the region $\mes > 5.2~\gevcc$
and $|\Delta E| < 0.25~\gev$. For an integrated luminosity
of $56.4~\invfb$, we obtain the preliminary branching
fractions
\begin{equation}
\BR(B^0 \ra K \ell^+ \ell^-) = 8.4 ^{+3.0 +1.0}_{-2.4 - 1.8} \times 10^{-7},
\label{KllBR}
\end{equation}
\begin{equation}
\BR(B^0 \ra K^* \ell^+ \ell^-) < 35 \times 10^{-7} (90\% C.L.),
\label{KstarllBR}
\end{equation}
where the first result represents an observation at the 5 sigma level
(statistical errors only).
Figure~\ref{fig:KllPlots} shows the projections from the likelihood fits
onto $\mes$ for the signal region in $\Delta E$.
\section{Conclusions}
We have shown a selection of results from the \babar\ experiment 
based on up to $56.4~\invfb$ collected at the \FourS\ resonance.
We have made the following observations:
\begin{itemize}
\item $B\ \ra DK$. We have measured
the ratio of the branching fractions for $B^- \ra D^0 K^-$ and
$B^- \ra D^0 \pi^-$, as well as the \CP\ asymmetry for the \CP-even
mode $B^- \ra D^0_+ K^-$, which is the first step towards measuring $\gamma$.
\item $B \ra D^{(*)} D^{(*)}$ decays, which can be used to measure $\beta$, 
have been fully reconstructed. We have a first
measurement of the \CP-odd content of these decays.
\item We observe $B^+ \ra \pi^+ \pi^0$ for the first time, which, with
other two body charmless modes, can be used to extract the angle $\alpha$.
\item Three-body charmless $B$ decays. Significant signals have been observed
for $B^+ \ra K^+ \pi^- \pi^+$ and $B^+ \ra K^+ K^- K^+$. A Dalitz plot analysis
of these decays could give us information about the angle $\gamma$.
\item We have made observations of the decays 
$B^+ \ra \phi K^+$ and $B^0 \ra \phi K^0$, and we also
confirm the rather large branching fractions of $\eta K^*$ and $\eta' K$
first seen by CLEO, which presents a theoretical challenge.
\item Radiative penguin modes. We observe a signal for $B^0 \ra K \ell^+ \ell^-$
for the first time.
\end{itemize}

Many of the results are approaching the level of predictions from the Standard
Model. We observe no direct \CP\ violation in several decays,
which could indicate that (the differences between) strong phases are small.
We can expect many more fruitful searches and improvements
to existing measurements in the near future.


\begin{thebibliography}{99}
\bibitem{babardet} \babar\ Collaboration, B.\ Aubert \etal\ 
``The \babar\ Detector'', \nimBaseB\ A {\bf 479}, 1 (2002), \hepex{0105044}.

\bibitem{Argus} ARGUS Collaboration, H.\ Albrecht \etal Z.\ Phys.\ C {\bf 48},
543 (1990).

\bibitem{gamma1} M.\ Gronau and D.\ Wyler, Phys.\ Lett.\ B {\bf 265}, 172 (1991);\\
M.\ Gronau and D.\ London, Phys.\ Lett.\ B {\bf 253}, 483 (1991).

\bibitem{gamma2} D.\ Atwood, I.\ Dunietz and A.\ Soni, Phys.\ Rev.\ Lett.
{\bf 78}, 3257 (1997).

\bibitem{pdg} Particle Data group, D.\ E.\ Groom \etal, Eur.\ Phys.\ Jour.\
C {\bf 15}, 1 (2000).

\bibitem{FoxWolfram} G.\ C.\ Fox and S.\ Wolfram, Phys.\ Rev.\ Lett.\ {\bf 41}, 
1581 (1978).

\bibitem{CLEODKR} M.\ Athanas \etal\ Phys.\ Rev.\ Lett. {\bf 80}, 5493 (1998),
hep-ex/9802023.

\bibitem{BELLEDKR} BELLE Collaboration, K.\ Abe \etal\ Phys.\ Rev.\ Lett. {\bf 87},
111801 (2001), hep-ex/0104051.

\bibitem{Dstar1} 
Y.\ Grossman and M.\ Worah, Phys.\ Lett.\ B {\bf 395}, 241 (1997);\\
R.\ Fleischer, Int.\ Jour.\ Mod.\ Phys.\ A {\bf 12}, 2459 (1997).

\bibitem{JpsiKs} \babar\ Collaboration, B.\ Aubert \etal\
SLAC-PUB-9293, hep-ex/0207042 (2002), submitted to Phys.\ Rev.\ Lett.

\bibitem{Dstar2} I.\ Dunietz \etal\ Phys.\ Rev.\ D {\bf 43}, 2193 (1991).

\bibitem{twobody}
M.\ Gronau and J.\ Rosner, 
Phys.\ Rev.\ Lett. {\bf 76}, 1200 (1996), hep-ph/9510363;\\
R.\ Fleischer and T.\ Mannel, Phys.\ Lett.\ B {\bf 397}, 269 (1997), hep-ph/9610357;

\bibitem{alpha} M.\ Gronau and D.\ London, Phys.\ Rev.\ Lett. {\bf 65}, 3381 (1990).

\bibitem{alphaBound} Y.\ Grossman and H.\ Quinn, Phys.\ Rev.\ D {\bf 58},
017504 (1998).
 
\bibitem{gamma3}
M.\ Neubert, Nucl.\ Phys.\ Proc.\ Suppl. {\bf 86}, 477 (2000), hep-ph/9909564.

\bibitem{hhref} \babar\ Collaboration, B.\ Aubert \etal\ Phys,\ Rev.\ Lett. {\bf 87},
151802 (2001), hep-ex/0105061.

\bibitem{sphericity} S.\ L.\ Wu, Phys.\ Rep.\ {\bf 107}, 59 (1984).

\bibitem{B3body} 
R.\ E.\ Blanco, C.\ Gobel, R.\ Mendez-Galain, Phys.\ Rev.\ Lett.
{\bf 86}, 2720 (2001), hep-ph/0007105; \\
S.\ Fajfer, R.\ J.\ Oakes, T.\ N.\ Pham,
Phys.\ Lett.\ B {\bf 539}, 67 (2002), hep-ph/0203072.

\bibitem{B3bodyBELLE} BELLE Collaboration, K.\ Abe \etal\ 
Phys.\ Rev.\ D {\bf 65}, 092005 (2002), hep-ex/0201007.

\bibitem{phiKQCD} H-Y.\ Cheng, Talk given at International
Europhysics Conference on High-Energy Physics (HEP 2001), Budapest,
Hungary, 12-18 Jul 2001, hep-ph/0110026;\\
S.\ Mishima, Phys.\ Lett.\ B {\bf 521}, 252 (2001), hep-ph/0107206.

\bibitem{phiKBELLE} BELLE Collaboration, K.\ Abe \etal\
KEK-PREPRINT-2001-74, BELLE-CONF-0113 (2001).

\bibitem{phiKCLEO} CLEO Collaboration, R.\ A.\ Briere \etal\
Phys.\ Rev.\ Lett. {\bf 86}, 3718 (2001), hep-ex/0101032.

\bibitem{etaCLEO} CLEO Collboration, S.\ J.\ Richichi \etal\
Phys.\ Rev.\ Lett. {\bf 85}, 520 (2000), hep-ex/9912059.

\bibitem{etaTheory} P.\ Ko, Talk given at $4^{\mbox{th}}$ International
Workshop on Particle Physics Phenomenology, Kaohsiung, Taiwan, China, 
18-21 Jun 1998, hep-ph/9810300.

\bibitem{Kstargamma} 
H.\ H.\ Asatrian, H.\ M.\ Asatrian, D.\ Wyler, Phys.\ Lett. B {\bf 470},
223 (1999), hep-ph/9905412.

\bibitem{KllTheory}
A.\ Ali \etal Phys.\ Rev.\ D {\bf 61}, 074024 (2000), hep-ph/9910221.

\end{thebibliography}
\end{document}